\documentclass[sigconf]{acmart}
\usepackage{graphicx}
\usepackage{hyperref}
\usepackage{enumitem}
\usepackage{lipsum}
\usepackage{multirow}

\setcopyright{acmlicensed}
\copyrightyear{2024}

\acmYear{2024}
\acmDOI{XXXXXXX.XXXXXXX}

\acmConference[GeoSim'24]{}{October 29, 2024}{Atlanta, GA}
\acmISBN{978-1-4503-XXXX-X/18/06}

\newcommand{\refsec}[1]{Section \ref{#1}}

\newcommand{\refequ}[1]{Equation \ref{#1}}

\newcommand*{\GitHubProposedApproach}[1]{\url{https://github.com/onspatial/geolife-plus#1}}

\renewcommand\footnotetextcopyrightpermission[1]{}

\begin{document}

\title{GeoLife+: Large-Scale Simulated Trajectory Datasets Calibrated to the GeoLife Dataset}

\author{Hossein Amiri}
\email{hossein.amiri@emory.edu}
\affiliation{%
    \institution{Emory University, Atlanta, USA}
    \city{}
    \country{}
}

\author{Richard Yang}
\email{richard.yang@emory.edu}
\affiliation{%
    \institution{Emory University, Atlanta, USA}
    \city{}
    \country{}
}
\author{Andreas Z{\"u}fle}
\email{azufle@emory.edu}
\affiliation{%
    \institution{Emory University, Atlanta, USA}
    \city{}
    \country{}
}

\renewcommand{\shortauthors}{Amiri, et al.}
\begin{abstract}
    Analyzing individual human trajectory data helps our understanding of human mobility and finds many commercial and academic applications. There are two main approaches to accessing trajectory data for research: one involves using real-world datasets like GeoLife, while the other employs simulations to synthesize data. Real-world data provides insights from real human activities, but such data is generally sparse due to voluntary participation. Conversely, simulated data can be more comprehensive but may capture unrealistic human behavior. In this Data and Resource paper, we combine the benefit of both by leveraging the statistical features of real-world data and the comprehensiveness of simulated data. Specifically, we extract features from the real-world GeoLife dataset such as the average number of individual daily trips, average radius of gyration, and maximum and minimum trip distances. We calibrate the Pattern of Life Simulation, a realistic simulation of human mobility, to reproduce these features. Therefore, we use a genetic algorithm to calibrate the parameters of the simulation to mimic the GeoLife features. For this calibration, we simulated numerous random simulation settings, measured the similarity of generated trajectories to GeoLife, and iteratively (over many generations) combined parameter settings of trajectory datasets most similar to GeoLife. Using the calibrated simulation, we simulate large trajectory datasets that we call GeoLife$^+$, where $^+$ denotes the Kleene Plus, indicating unlimited replication with at least one occurrence. We provide simulated GeoLife$^+$ data with 182, 1k, and 5k over 5 years, 10k, and 50k over a year and 100k users over 6 months of simulation lifetime.

\end{abstract}

\ccsdesc[500]{Information systems~Geographic information systems \vspace{-0.03cm}}
\ccsdesc[500]{Information systems~Location based services\vspace{-0.2cm} }
\keywords{GeoLife, Patern of Life, Simulation, Trajectory Datasets }
\vspace{-0.3cm}
\maketitle

\vspace{-0.3cm}
\section{Introduction}
\begin{figure}[t]
    \centering
    \includegraphics[trim = 2cm 2cm 2cm 4cm , clip,width=1\linewidth]{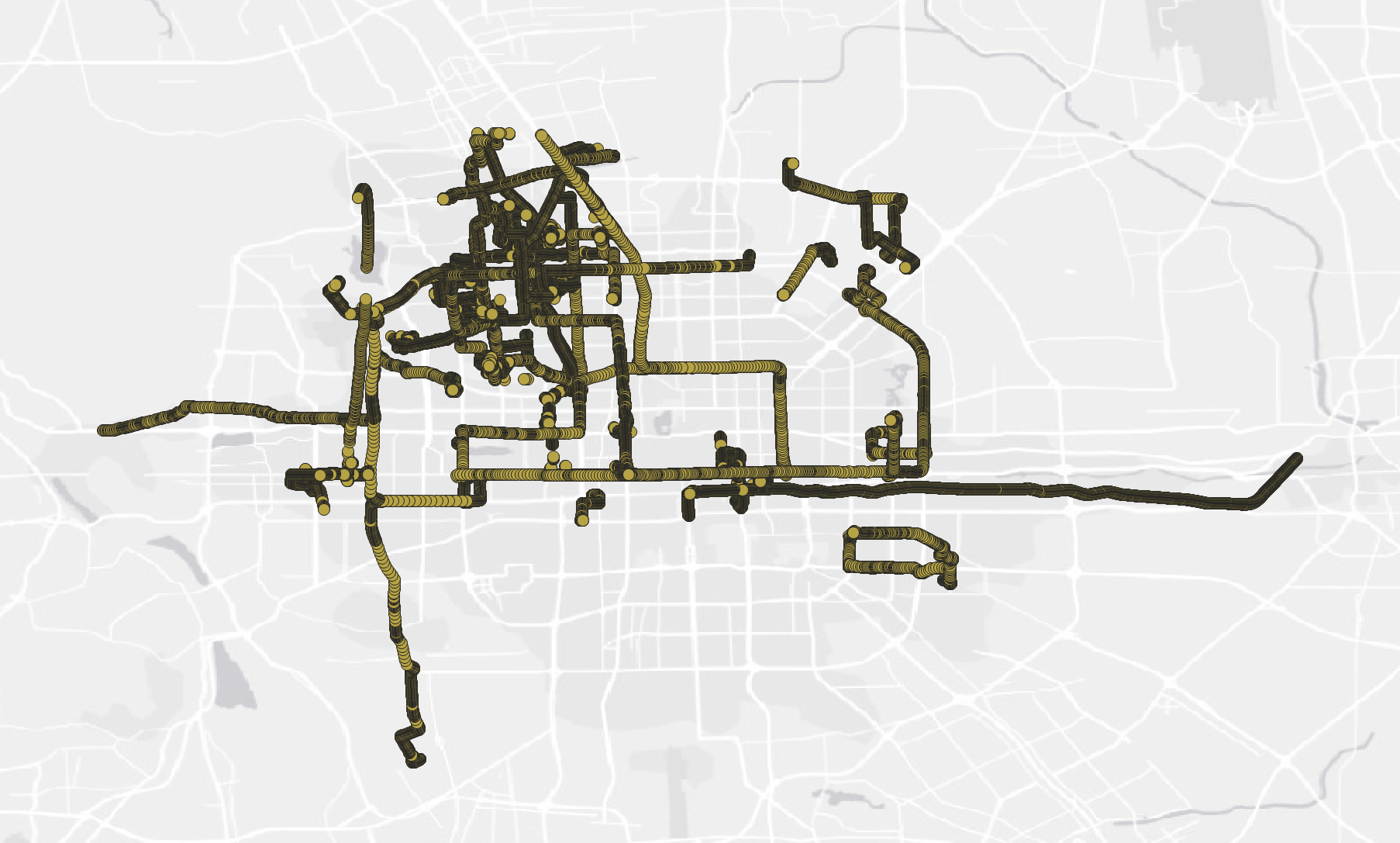}
    \vspace{-0.35689cm}
    \caption{ GeoLife dataset visualization for the busiest day in Beijing, China \vspace{-0.1cm} }
    \label{fig:geolife_gps}
    \vspace{-0.1cm}
\end{figure}

Trajectory data \cite{amiri2023massive,amiri2024pattern} is essential for analyzing human behavior \cite{toch2019analyzing,zhu2024generic} and mobility analysis \cite{mokbel2018mobility}, traffic analysis \cite{chen2022lane},  providing insights that are valuable for applications like outlier detection \cite{zhang2023large,zhang2024transeferable} and urban planning \cite{isaacman2012human}. Analyzing location data derived from real human mobility can lead to better-informed decisions, but real-world data is difficult to obtain due to location privacy~\cite{xiao2015protecting,khoshgozaran2011location,krumm2009survey}. Consequently, openly available trajectory datasets are collected by usually small numbers of volunteers. The largest and most commonly used trajectory dataset is the GeoLife Dataset~\cite{zheng2011geolife} which captures 182 individuals in Beijing, China over more than five years. However, in GeoLife only 45 users have more than 100 staypoints, indicating that the dataset is sparse and patterns focuses primarily on its major users. In addition, the maximum number of GeoLife users active on any given day is only 25 in Beijing. Figure~\ref{fig:geolife_gps} shows the trajectory of these 25 users on one of the busiest GeoLife days. We see that these trajectories are distributed very sparsely over the Beijing area. Given the population of Beijing of more than 20 million people, it appears impossible to find representative patterns of human mobility from no more than 25 users. This makes it very challenging to employ GeoLife for the many applications.

Due to these limitations, researchers often employ simulated trajectory data: it offers a richer and cleaner dataset without privacy concerns. However, many existing trajectory simulations have agents travel to uniformly random destinations~\cite{duntgen2009berlinmod,brinkhoff2002framework} or use parametric distributions to generate trip destinations~\cite{levandoski2012lars,armenatzoglou2013general}. While such trajectories are useful for evaluating database storage and retrieval solutions for trajectories, the randomness of mobility does not allow to improve our understanding of human mobility. Towards a more realistic simulation of human mobility, the Pattern of Life Simulation (POL) \cite{kim2020location,zufle2023urban} simulates the physiological, safety, love, and esteem needs of individual agents to create trips with a purpose such as going home to sleep, going to work to make money, or going to a restaurant to eat.

In this work, we leverage the Patterns of Life simulation to generate a set of datasets that are statistically similar to GeoLife's active user data. This approach allows us to 1) maintain the socially plausible agent behavior inherited from the Patterns of Life simulation, 2) while having descriptive statistics of the trajectory dataset closely reflecting the original GeoLife data, and 3) having a substantially higher density of data points compared to GeoLife. To identify the optimal parameter configuration for the Patterns of Life Simulation to generate GeoLife-like data, we developed a genetic algorithm that initially explores random simulation parameters and iteratively ``crosses'' parameter settings having high similarity to GeoLife.  Our primary contributions in this Data \& Resource paper is as follows:


\begin{itemize}[noitemsep,
        nosep,
        leftmargin=10pt,
        labelsep=2pt,
        itemindent=0pt]
    \item We provide multiple trajectory datasets  collected over 5 years of simulation periods for 182, 1k, and 5k agents,  1 year for 10k and 50k agents, and 6 months for 100k agents. These datasets sized at hundreds of gigabyte of trajectories are available at OSF link provided on \GitHubProposedApproach{}
    \item A genetic algorithm used to calibrate parameters based on the GeoLife dataset, with instructions on how to apply various statistical methods to generate additional datasets. The source code can be accesses on the GitHub repository.
    \item The simulation parameters resulting from this geometric algorithm and used for the generation of the aforementioned datasets. Together with our documentation to re-run the simulation, this allows users to re-generate the data without having to download large datasets and allows users to create even larger datasets having more agent or having longer simulation periods. Parameter configurations and simulation documentation are available on the GitHub repository.
\end{itemize}

The remainder of the paper is organized as follows:
In \refsec{sec:prior_datasets}, we review the existing trajectory datasets, both real and synthetic.
In \refsec{sec:methodology}, we explain our methodology, including data generation, the genetic algorithm, and the steps to create the dataset.
In \refsec{sec:dataset}, we describe and analyze the generated data.
In \refsec{sec:regeneration} regeneration process is described. 
Finally, we summarize our findings and draw conclusions in \refsec{sec:conclusion}.

\vspace{-0.3cm}
\section{Existing Trajectory Datasets}
\label{sec:prior_datasets}

Trajectory datasets have been widely studied previously. Trajectories can be recorded for humans, animals, vehicles, etc., using sensors in real-world settings \cite{zheng2011geolife, yabe2024yjmob100k} or synthesizing the data using simulations \cite{amiri2023massive}, generative models \cite{zhang2022factorized} and dataset enhancement \cite{hu2022processing,zhu2024generic,zhu2024synmob}. In this section, we describe the most commonly used real-world and synthesized individual human trajectories.

%
\subsection{Real World Human Trajectory Data}
The GeoLife GPS trajectory dataset \cite{zheng2011geolife}, collected by Microsoft Research Asia, is one of the most commonly used real-world trajectory datasets. It encompasses trajectories of ~180 users in Beijing, China, over a period of more than four years (from April 2007 to October 2011). The dataset captures a wide range of movements, not only routine activities like commuting to work or returning home, but also leisure activities such as shopping, sightseeing, dining, hiking, and cycling. Although the dataset is of high quality and fidelity, its relatively small size of only 180 users makes it challenging to infer broad mobility patterns, particularly in a large urban area like Beijing.
A large-scale real-world human mobility dataset was introduced recently in \cite{yabe2024yjmob100k}. The data were collected from mobile phones on a metropolitan scale and encompasses observations from 100,000 individuals over a period of 75 days. 
Data collection was conducted with user consent, and to ensure privacy, all data was anonymized. To further protect privacy, the data is structured into a spatial grid having spatial cells measuring 500 meters by 500 meters, and data is recorded at 30-minute intervals. This relatively low spatial granularity allows understanding high-level human mobility, but can't be used to infer visit patterns at individual places of interest. 
In addition, numerous trajectory datasets do not pertain to individual humans, but to vehicles such as taxi trajectories in Beijing, China~\cite{yuan2010t} and San Francisco~\cite{PSG09}, and bus trajectories in Rio de Janeiro, Brazil~\cite{DC18}. While such datasets can be used to understand traffic patterns based on travel speeds, it is difficult to infer human mobility patterns from such data, as a taxi may capture a different and independent human passenger each trip. 
\subsection{Synthetically Generated Trajectories}

Deep generative models have been proposed recently in \cite{zhang2022factorized} to addressing the scarcity of large real datasets. Their "End-to-End Trajectory Generation with Spatiotemporal Validity Constraints" (EETG) framework significantly improves trajectory synthesis, showcasing the similarity of generated trajectories to real-world trajectories. 
%
Towards dataset enhancement, a generic optimization-based approach was introduced recently~\cite{zhu2024generic}, which utilizes both position and velocity data as a baseline to enhance driving trajectory data.
In \cite{zhu2024synmob} a diffusion model is utilized to synthesize the spatial-temporal behavior of the original dataset accurately. This model learns complex spatial-temporal motion patterns and emulates the geographical distribution and statistical properties of real-world trajectories. However, the goal of these generated datasets is to mimic kinematic trajectory features rather than creating  human mobility patterns.


The Patterns of Life Simulation \cite{kim2020location} allows generating city-level human mobility data. The simulation uses data from OpenStreetMap to model agents moving between various locations like home, work, restaurants, and recreational sites. Agents' activities are guided by Maslowian needs \cite{maslow1943theory}, including basic physiological needs, financial needs, and social needs, which drive their decisions and interactions. A detailed description of the simulation can be found in~\cite{zufle2023urban}.
In \cite{amiri2023massive}, the Patterns of Life Simulation was used to generate a massive trajectory dataset. 
The dataset comprises over 1.5 terabytes of simulated data, which includes more than 22,360,320,024 trajectory locations, over 423,609,129 check-ins, and more than 1,736,701,154 social links. The Patterns of Life Simulation has a very large number of parameters to define the agents' needs and consequently, their behavior. 
While the default parameters used for this dataset are realistic, we do not have a clear understanding how these parameters may differ at different places around the world. In this paper, we fill this gap by calibrating the Patterns of Life Simulation to GeoLife to find near-optimal parameter settings to reproduce features observed in GeoLife data for Beijing.

\vspace{-0.3cm}
\section{Simulation Calibration}
\label{sec:methodology}
In this section, we explain our approach to generate the GeoLife$^*$ dataset. Specifically, we describe the similarity function we calculated to assess the similarity of the original GeoLife dataset and the generated datasets in Section~\ref{sec:similarity}. Based on this similarity function, we describe the genetic algorithm developed to calibrate simulation parameters in Section~\ref{sec:genetic}, and we provide the results of this simulation calibration in Section~\ref{sec:calliresults}.

\subsection{Trajectory Data Similarity}\label{sec:similarity}
As GeoLife trajectories travel outside of Beijing (and even of China), we restricted the trajectories to the Beijing area with the bounding box of [39.748, 116, 165, 40.038, 116.628]. Within this region, we first extracted staypoints using the algorithm described in~\cite{li2008mining} to find the places that users visit and delineate the trips connecting these places. For this calibration, we only use GeoLife users at least 100 staypoints. This filter yielded approximately 12,000 staypoints for 45 users. For these users, we computed the average distance per trip (ADT) as $3692.13m$, the average distance per agent per day(ADA) as $4474.59m$, the maximum trip distance (MXD) as $30262m$ and the median trip distance (MDD) as $3349.75m$. 
We applied the same metrics to the simulated data. For each set of simulated data, we used the formula in \refequ{equ:score} to score the similarity of the simulated check-ins to the GeoLife check-ins. A score closer to 1 indicates greater similarity between the simulated and GeoLife check-ins.


\begin{equation}
    \label{equ:score}
    \text{$Similarity(G,P)$} = 1 - \frac{1}{|\text{$M$}|} \sum_{k \in \text{$M$}} \frac{\left| k(P) - k(G) \right|}{k(G)}
\end{equation}

where $G$ is the GeoLife dataset, $P$ is a set of simulated trajectories, ${M} = \{ ADT, ADA, MXD, MDD \}$ represents the set of metrics, $k(P)$ represents the results of metric $k$ on the Patterns of Life dataset, and $k(P)$ is the result of metric $k$ on GeoLife.
\subsection{Genetic Algorithm for Calibration}\label{sec:genetic}
Initially, we manually identified 63 simulation parameters we deemed relevant in defining the agents' behavior. These parameters included factors such as the number of agents' interests, the maximum allowed rental salary ratio, and the agents' walking speed. The full list of all parameters used for calibration can be found on our GitHub repository. Our goal was to find values for these parameters that yield a simulation of Beijing that most closely mimics the trajectory metrics observed on GeoLife. 

%
%
%
%
We initialize the algorithm using $n$ (``layer size'') simulation runs using randomly chosen (within manually chosen ranges deemed plausible) parameter values. For each trajectory dataset by a simulation, we used Equation~\ref{equ:score} to find parameter settings yielding the top simulations most similar to GeoLife. These simulations initialized the genetic algorithm in which these parameter settings (``parents'') were combined into in five different ways to create new parameter settings (``child'') by choosing, for each parameter, at random one of the following: 1) the maximum parameter value of the selected parents, 2) the minimum parameter value of the selected parents, 3) the mean parameter value of the selected parents, 4) random combinations of the values from the selected parents, or 5) a new parameter value chosen at random (``mutation''). This random combination of attributes was repeated until $n$ children were generated. 
For each of these $n$ children, a corresponding simulation (with the selected parameter setting) was run. Then, Equation~\ref{equ:score} was again used to find the children yielding trajectory data most similar to GeoLife. Using these children, this process is repeated to create a new layers (``generation'') of simulations. This process of creating new generations of simulations is repeated indefinitely until manually stopped. The parameter settings (across all simulated generations) yielding the most similar trajectories to GeoLife is the chosen as the result of the calibration step. The source code of this genetic algorithm can be found on \GitHubProposedApproach{}.
\subsection{GeoLife Calibration Results}\label{sec:calliresults}
Running the genetic algorithm with layer sizes of 8, 32, 64 and 128 yielded 10, 15, 32, and 78 configurations, respectively, that achieved similarity scores (using Equation~\ref{equ:score}) higher than 0.8. We retained the top 10 configurations for subsequent dataset scaling. These configurations are accessible at \GitHubProposedApproach{/blob/main/restults/params.top.json}. We utilize the \textit{params.top.json} file in the subsequent phase to generate datasets and evaluate their similarities to the original GeoLife dataset across varying numbers of agents and simulation setups.

\vspace{-0.3cm}
\section{Dataset Description}
\label{sec:dataset}
\begin{figure}
    \centering
    \includegraphics[trim = 2cm 2cm 2cm 2cm, clip, width=1\linewidth]{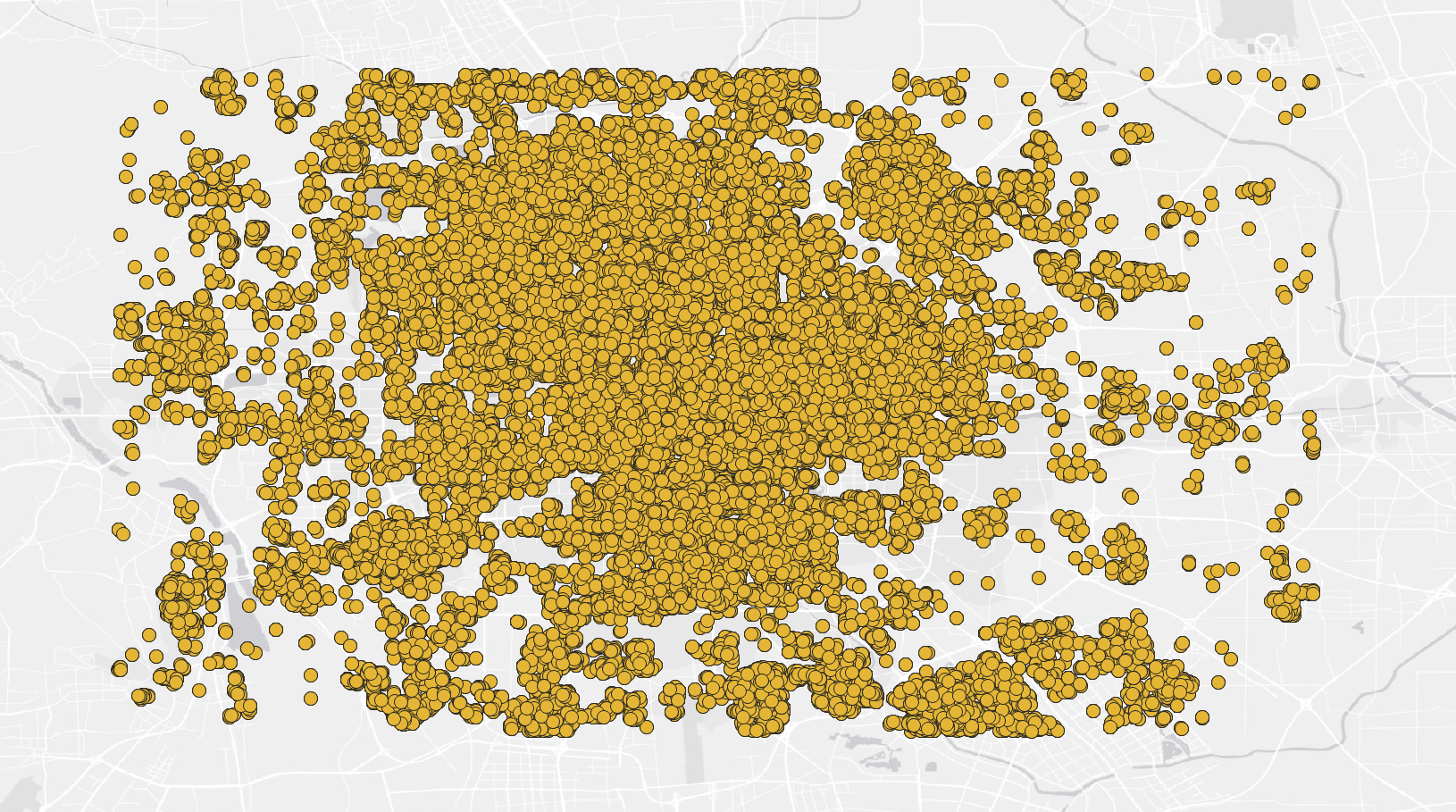}
    \caption{Staypoints of a simulation day with 50k agents.}
    \label{fig:pol_checkins}
\end{figure}

Figure~\ref{fig:pol_checkins} shows the location of staypoints observed in a single day of GeoLife$^*$ with 50,000 agents. This figure omits the trajectories of agents between staypoint. We observe that agents visit a location in Beijing across the entire city, giving a much more representative coverage than a day of GeoLife shown in Figure~\ref{fig:geolife_gps}.

The datasets generated in this study, as detailed in Table~\ref{tab:dataset_specification} and Table~\ref{tab:dataset_similarity}, capture a range of simulated scenarios based on the original GeoLife dataset, scaled to various agent populations and time periods. 
Table~\ref{tab:dataset_specification} shows the size (data size, \# staypoints, \# GPS updates) for our generated GeoLife$^*$ datasets and the original GeoLife dataset. Running GeoLife$^*$ with the same number of agents and duration (182 users, 5 Years), we observe that GeoLife$^*$ is approximately 100 times larger. This is because in GeoLife$^*$ all 182 agents are fully observed every day, whereas users in GeoLife become inactive for long periods. The number of GPS Updates in GeoLife$^*$ is only about five times larger than GeoLife. That's because GeoLife$^*$ uses five-minute frequency location updates (to minimize storage cost), whereas GeoLife users are observed every 1–5 seconds. We note that the sampling frequency of the Patterns of Life simulation can be changed as a parameter. But we see that by simulating 1k agents for five simulation years, the dataset already grows to 180GB even at a $1/300$Hz sampling frequency.

Table~\ref{tab:dataset_similarity} additional shows the metrics we used to calibrate the simulation for each of the generated datasets and the resulting similarity score defined in Equation~\ref{equ:score}. 
As the number of agents increases, we observe that the average distance per agent per day (ADA) decreases. We explain this due to the simulation creating more recreational sites to accommodate the large number of agents. Thus, agent are able to find recreational sites closer to their home. At the same time, the agent's work location (which is usually further than recreational sites and, thus, defined the agents' radii of gyration) remains similar as the number of agents increases.

Additionally, we present the dataset titled "59k-5yrs," a carefully curated compilation of the top 300 most similar generated datasets from the "182-5yrs" series. These datasets have been selected based on their high similarity, resulted in a total similarity score of 0.74 to the original Geolife dataset. The "59k-5yrs" dataset, with a total size of 41GB, contains approximately 575 million staypoints.

\section{Data Sharing and Re-Generation}
\label{sec:regeneration}
We're sharing the datasets sized smaller than 100GB on \texttt{OSF.io}. Links to these datasets can be found on our GitHub repository. For datasets larger than 100GB, our GitHub documentation provides instructions how to run the simulation and locally re-generate the data. This will allow researchers to generate even larger (in terms of the number of agents or simulation duration) datasets. For this purpose, Table~\ref{tab:dataset_specification} also shows the wall-close time the simulation took to run on a compact desktop machine having a 2.40Ghz i5-1135G7 processor with eight cores and 16GB of main memory running Linux Fedora.

\begin{table}[t]
    \centering

    \begin{tabular}{|l|c|c|c|c|c|}
        \hline
        & \multicolumn{2}{c|}{Staypoints} & \multicolumn{2}{c|}{GPS  Updates}  &  \\
         \hline
        Dataset  & Size  & Number  & Size& Number & Time \\
        \hline
        GeoLife & 1.7 MB &  19K        & 1.7GB & 28M   &   N/A \\
        \hline
        182-5yrs & 178 MB &  ~2.2M        & ~30GB & ~120M   &   1.72h \\
        \hline
        1k-5yrs  & ~0.6GB   & ~8M       & ~180GB   & ~600M  & 8.2h   \\
        \hline
        5k-5yrs  & 3.1GB   & ~39M       & NP  & NP   &  18.46h\\
        \hline
        10k-1yr  & 1.2GB  & 15.5M           & NP  & NP  &   11.75h \\
        \hline
        50k-1yr  & 16GB  & 200M        & NP & NP  &  127.45h \\
        \hline
        100k-6mo & 16GB & 200M       & NP & NP   & 139.42h \\
        \hline
        59k-5yrs  & 41GB   & ~575M       & NP  & NP   &  NA\\
        \hline
    \end{tabular}
    \caption{Specification of GeoLife scaled datasets. Each dataset name follows the format [\#agents]-[simulated time]. For example, "1k-5yrs" refers to a dataset where 1,000 agents were simulated for 5 years.  (K: Thousnd, M: Million, B: Billion, T: Trillion). NP: Not provided because the size was very large and logging the data was storage and time-consuming }
    \label{tab:dataset_specification}
\end{table}



\begin{table}[t]
    \centering

    \begin{tabular}{|l|c|c|c|c|c|}
        \hline
        Dataset  & ADT     & ADA     & MXD     & MDD    & Score \\
        \hline
        GeoLife & 3692.13 & 4474.59 & 30262 & 3349.75 & 1  \\
        \hline
        182-5yrs & 4217.12 & 4100.51 & 29013.0 & 3346.0 &  0.93  \\
        \hline
        1k-5yrs  & 3557.37       & 3445.18     & 34365.0      & 2815.0      & 0.85  \\
        \hline
        5k-5yrs  & 1217.29       & 1209.29       & 29202.0       & 885.0      & 0.45  \\
        \hline
        10k-1yr  & 909.34       & 910.11      & 29679.0       & 667.0     & 0.40 \\
        \hline
        50k-1yr  & 475.63       & 473.67       & 31978.0       & 369.0      & 0.32  \\
        \hline
        100k-6mo & 427.96       & 425.79       & 38612.0       & 326.0      & 0.25  \\
        \hline
        59k-5yrs  & 5073.91       & 4884.97      & 39886.0       & 4165.0     & 0.74  \\
        \hline
    \end{tabular}
    \caption{Geo-statistics of the generated datasets. \vspace{-0.5cm}}
    \label{tab:dataset_similarity}
\end{table}




\vspace{-0.3cm}
\section{Conclusions}
\label{sec:conclusion}
In this paper, we have presented a novel approach to generate large-scale synthetic geospatial datasets by simulating the pattern of life of individuals with the consideration of real-world constraints. We have demonstrated the effectiveness of our approach by generating a set of synthetic datasets based on the GeoLife dataset. We have shown that the generated datasets exhibit similar statistical properties to the original dataset, and can be used for various geospatial data analysis tasks. We have also provided detailed instructions on how to reproduce the generated datasets, and have made the code and data available on GitHub. With the provided datasets and code, researchers can easily generate large-scale synthetic geospatial datasets for their research purposes and evaluate the performance of their algorithms on realistic data.

\bibliographystyle{ACM-Reference-Format}
\bibliography{main}


\begin{thebibliography}{28}


\ifx \showCODEN    \undefined \def \showCODEN     #1{\unskip}     \fi
\ifx \showDOI      \undefined \def \showDOI       #1{#1}\fi
\ifx \showISBNx    \undefined \def \showISBNx     #1{\unskip}     \fi
\ifx \showISBNxiii \undefined \def \showISBNxiii  #1{\unskip}     \fi
\ifx \showISSN     \undefined \def \showISSN      #1{\unskip}     \fi
\ifx \showLCCN     \undefined \def \showLCCN      #1{\unskip}     \fi
\ifx \shownote     \undefined \def \shownote      #1{#1}          \fi
\ifx \showarticletitle \undefined \def \showarticletitle #1{#1}   \fi
\ifx \showURL      \undefined \def \showURL       {\relax}        \fi
\providecommand\bibfield[2]{#2}
\providecommand\bibinfo[2]{#2}
\providecommand\natexlab[1]{#1}
\providecommand\showeprint[2][]{arXiv:#2}

\bibitem[Amiri et~al\mbox{.}(2024)]%
        {amiri2024pattern}
\bibfield{author}{\bibinfo{person}{Hossein Amiri}, \bibinfo{person}{Will Kohn}, {et~al\mbox{.}}} \bibinfo{year}{2024}\natexlab{}.
\newblock \showarticletitle{The Patterns of Life Human Mobility Simulation}. In \bibinfo{booktitle}{\emph{SIGSPATIAL'24 (To Appear, Demonstration Track)}}.
\newblock


\bibitem[Amiri et~al\mbox{.}(2023)]%
        {amiri2023massive}
\bibfield{author}{\bibinfo{person}{Hossein Amiri}, \bibinfo{person}{Shiyang Ruan}, \bibinfo{person}{Joon-Seok Kim}, {and} \bibinfo{person}{other}.} \bibinfo{year}{2023}\natexlab{}.
\newblock \showarticletitle{Massive Trajectory Data Based on Patterns of Life}. In \bibinfo{booktitle}{\emph{SIGSPATIAL '23}}. \bibinfo{pages}{1--4}.
\newblock


\bibitem[Armenatzoglou et~al\mbox{.}(2013)]%
        {armenatzoglou2013general}
\bibfield{author}{\bibinfo{person}{Nikos Armenatzoglou}, \bibinfo{person}{Stavros Papadopoulos}, {and} \bibinfo{person}{Dimitris Papadias}.} \bibinfo{year}{2013}\natexlab{}.
\newblock \showarticletitle{A general framework for geo-social query processing}.
\newblock \bibinfo{journal}{\emph{Proc. of the VLDB Endowment}} \bibinfo{volume}{6}, \bibinfo{number}{10} (\bibinfo{year}{2013}), \bibinfo{pages}{913--924}.
\newblock


\bibitem[Brinkhoff(2002)]%
        {brinkhoff2002framework}
\bibfield{author}{\bibinfo{person}{Thomas Brinkhoff}.} \bibinfo{year}{2002}\natexlab{}.
\newblock \showarticletitle{A framework for generating network-based moving objects}.
\newblock \bibinfo{journal}{\emph{GeoInformatica}} \bibinfo{volume}{6}, \bibinfo{number}{2} (\bibinfo{year}{2002}), \bibinfo{pages}{153--180}.
\newblock


\bibitem[Chen et~al\mbox{.}(2022)]%
        {chen2022lane}
\bibfield{author}{\bibinfo{person}{Wenqiang Chen}, \bibinfo{person}{Tao Wang}, {et~al\mbox{.}}} \bibinfo{year}{2022}\natexlab{}.
\newblock \showarticletitle{Lane-based Distance-Velocity model for evaluating pedestrian-vehicle interaction at non-signalized locations}.
\newblock \bibinfo{journal}{\emph{Accident Analysis \& Prevention}}  \bibinfo{volume}{176} (\bibinfo{year}{2022}), \bibinfo{pages}{106810}.
\newblock


\bibitem[Dias and Costa(2018)]%
        {DC18}
\bibfield{author}{\bibinfo{person}{Daniel Dias} {and} \bibinfo{person}{Luis Henrique Maciel~Kosmalski Costa}.} \bibinfo{year}{2018}\natexlab{}.
\newblock \bibinfo{title}{{CRAWDAD} dataset coppe-ufrj/RioBuses (v. 2018-03-19)}.
\newblock \bibinfo{howpublished}{Downloaded from \url{https://crawdad.org/coppe-ufrj/RioBuses/20180319}}.
\newblock


\bibitem[D{\"u}ntgen et~al\mbox{.}(2009)]%
        {duntgen2009berlinmod}
\bibfield{author}{\bibinfo{person}{Christian D{\"u}ntgen}, \bibinfo{person}{Thomas Behr}, {and} \bibinfo{person}{Ralf~Hartmut G{\"u}ting}.} \bibinfo{year}{2009}\natexlab{}.
\newblock \showarticletitle{BerlinMOD: a benchmark for moving object databases}.
\newblock \bibinfo{journal}{\emph{The VLDB Journal}}  \bibinfo{volume}{18} (\bibinfo{year}{2009}), \bibinfo{pages}{1335--1368}.
\newblock


\bibitem[Hu et~al\mbox{.}(2022)]%
        {hu2022processing}
\bibfield{author}{\bibinfo{person}{Xiangwang Hu}, \bibinfo{person}{Zuduo Zheng}, \bibinfo{person}{Danjue Chen}, {et~al\mbox{.}}} \bibinfo{year}{2022}\natexlab{}.
\newblock \showarticletitle{Processing, assessing, and enhancing the Waymo autonomous vehicle open dataset for driving behavior research}.
\newblock \bibinfo{journal}{\emph{Transportation Research Part C: Emerging Technologies}}  \bibinfo{volume}{134} (\bibinfo{year}{2022}), \bibinfo{pages}{103490}.
\newblock


\bibitem[Isaacman et~al\mbox{.}(2012)]%
        {isaacman2012human}
\bibfield{author}{\bibinfo{person}{Sibren Isaacman}, \bibinfo{person}{Richard Becker}, {et~al\mbox{.}}} \bibinfo{year}{2012}\natexlab{}.
\newblock \showarticletitle{Human mobility modeling at metropolitan scales}. In \bibinfo{booktitle}{\emph{Proceedings of the 10th international conference on Mobile systems, applications, and services}}. \bibinfo{pages}{239--252}.
\newblock


\bibitem[Khoshgozaran et~al\mbox{.}(2011)]%
        {khoshgozaran2011location}
\bibfield{author}{\bibinfo{person}{Ali Khoshgozaran}, \bibinfo{person}{Cyrus Shahabi}, {and} \bibinfo{person}{Houtan Shirani-Mehr}.} \bibinfo{year}{2011}\natexlab{}.
\newblock \showarticletitle{Location privacy: going beyond K-anonymity, cloaking and anonymizers}.
\newblock \bibinfo{journal}{\emph{Knowledge and Information Systems}}  \bibinfo{volume}{26} (\bibinfo{year}{2011}), \bibinfo{pages}{435--465}.
\newblock


\bibitem[Kim et~al\mbox{.}(2020)]%
        {kim2020location}
\bibfield{author}{\bibinfo{person}{Joon-Seok Kim}, \bibinfo{person}{Hyunjee Jin}, \bibinfo{person}{Hamdi Kavak}, {et~al\mbox{.}}} \bibinfo{year}{2020}\natexlab{}.
\newblock \showarticletitle{Location-based social network data generation based on patterns of life}. In \bibinfo{booktitle}{\emph{MDM}}. IEEE, \bibinfo{pages}{158--167}.
\newblock


\bibitem[Krumm(2009)]%
        {krumm2009survey}
\bibfield{author}{\bibinfo{person}{John Krumm}.} \bibinfo{year}{2009}\natexlab{}.
\newblock \showarticletitle{A survey of computational location privacy}.
\newblock \bibinfo{journal}{\emph{Personal and Ubiquitous Computing}}  \bibinfo{volume}{13} (\bibinfo{year}{2009}), \bibinfo{pages}{391--399}.
\newblock


\bibitem[Levandoski et~al\mbox{.}(2012)]%
        {levandoski2012lars}
\bibfield{author}{\bibinfo{person}{Justin~J Levandoski}, \bibinfo{person}{Mohamed Sarwat}, \bibinfo{person}{Ahmed Eldawy}, {and} \bibinfo{person}{Mohamed~F Mokbel}.} \bibinfo{year}{2012}\natexlab{}.
\newblock \showarticletitle{Lars: A location-aware recommender system}. In \bibinfo{booktitle}{\emph{ICDE}}. IEEE, \bibinfo{pages}{450--461}.
\newblock


\bibitem[Li et~al\mbox{.}(2008)]%
        {li2008mining}
\bibfield{author}{\bibinfo{person}{Quannan Li}, \bibinfo{person}{Yu Zheng}, \bibinfo{person}{Xing Xie}, {et~al\mbox{.}}} \bibinfo{year}{2008}\natexlab{}.
\newblock \showarticletitle{Mining user similarity based on location history}. In \bibinfo{booktitle}{\emph{Proceedings of the 16th ACM SIGSPATIAL international conference on Advances in geographic information systems}}. \bibinfo{pages}{1--10}.
\newblock


\bibitem[Maslow(1943)]%
        {maslow1943theory}
\bibfield{author}{\bibinfo{person}{Abraham~H Maslow}.} \bibinfo{year}{1943}\natexlab{}.
\newblock \showarticletitle{A theory of human motivation.}
\newblock \bibinfo{journal}{\emph{Psychological review}} \bibinfo{volume}{50}, \bibinfo{number}{4} (\bibinfo{year}{1943}), \bibinfo{pages}{370}.
\newblock


\bibitem[Mokbel et~al\mbox{.}(2024)]%
        {mokbel2018mobility}
\bibfield{author}{\bibinfo{person}{Mohamed Mokbel}, \bibinfo{person}{Mahmoud Sakr}, \bibinfo{person}{}, {et~al\mbox{.}}} \bibinfo{year}{2024}\natexlab{}.
\newblock \showarticletitle{Mobility Data Science: Perspectives and Challenges}.
\newblock \bibinfo{journal}{\emph{ACM Transactions on Spatial Algorithms and Systems}} (\bibinfo{year}{2024}).
\newblock


\bibitem[Piorkowski et~al\mbox{.}(2009)]%
        {PSG09}
\bibfield{author}{\bibinfo{person}{Michal Piorkowski}, \bibinfo{person}{Natasa Sarafijanovic-Djukic}, {and} \bibinfo{person}{Matthias Grossglauser}.} \bibinfo{year}{2009}\natexlab{}.
\newblock \bibinfo{title}{{CRAWDAD} dataset epfl/mobility (v. 2009-02-24)}.
\newblock \bibinfo{howpublished}{Downloaded from \url{https://crawdad.org/epfl/mobility/20090224}}.
\newblock


\bibitem[Toch et~al\mbox{.}(2019)]%
        {toch2019analyzing}
\bibfield{author}{\bibinfo{person}{Eran Toch}, \bibinfo{person}{Boaz Lerner}, \bibinfo{person}{Eyal Ben-Zion}, {and} \bibinfo{person}{Irad Ben-Gal}.} \bibinfo{year}{2019}\natexlab{}.
\newblock \showarticletitle{Analyzing large-scale human mobility data: a survey of machine learning methods and applications}.
\newblock \bibinfo{journal}{\emph{Knowledge and Information Systems}}  \bibinfo{volume}{58} (\bibinfo{year}{2019}), \bibinfo{pages}{501--523}.
\newblock


\bibitem[Xiao and Xiong(2015)]%
        {xiao2015protecting}
\bibfield{author}{\bibinfo{person}{Yonghui Xiao} {and} \bibinfo{person}{Li Xiong}.} \bibinfo{year}{2015}\natexlab{}.
\newblock \showarticletitle{Protecting locations with differential privacy under temporal correlations}. In \bibinfo{booktitle}{\emph{Proceedings of the 22nd ACM SIGSAC conference on computer and communications security}}. \bibinfo{pages}{1298--1309}.
\newblock


\bibitem[Yabe et~al\mbox{.}(2024)]%
        {yabe2024yjmob100k}
\bibfield{author}{\bibinfo{person}{Takahiro Yabe}, \bibinfo{person}{Kota Tsubouchi}, {et~al\mbox{.}}} \bibinfo{year}{2024}\natexlab{}.
\newblock \showarticletitle{YJMob100K: City-scale and longitudinal dataset of anonymized human mobility trajectories}.
\newblock \bibinfo{journal}{\emph{Scientific Data}} \bibinfo{volume}{11}, \bibinfo{number}{1} (\bibinfo{year}{2024}), \bibinfo{pages}{397}.
\newblock


\bibitem[Yuan et~al\mbox{.}(2010)]%
        {yuan2010t}
\bibfield{author}{\bibinfo{person}{Jing Yuan}, \bibinfo{person}{Yu Zheng}, \bibinfo{person}{Chengyang Zhang}, {et~al\mbox{.}}} \bibinfo{year}{2010}\natexlab{}.
\newblock \showarticletitle{T-drive: driving directions based on taxi trajectories}. In \bibinfo{booktitle}{\emph{Proceedings of the 18th SIGSPATIAL International conference on advances in geographic information systems}}. \bibinfo{pages}{99--108}.
\newblock


\bibitem[Zhang et~al\mbox{.}(2022)]%
        {zhang2022factorized}
\bibfield{author}{\bibinfo{person}{Liming Zhang}, \bibinfo{person}{Liang Zhao}, {and} \bibinfo{person}{Dieter Pfoser}.} \bibinfo{year}{2022}\natexlab{}.
\newblock \showarticletitle{Factorized deep generative models for end-to-end trajectory generation with spatiotemporal validity constraints}. In \bibinfo{booktitle}{\emph{SIGSPATIAL '22}}. \bibinfo{pages}{1--12}.
\newblock


\bibitem[Zhang et~al\mbox{.}(2024)]%
        {zhang2024transeferable}
\bibfield{author}{\bibinfo{person}{Zheng Zhang}, \bibinfo{person}{Hossein Amiri}, \bibinfo{person}{}, {et~al\mbox{.}}} \bibinfo{year}{2024}\natexlab{}.
\newblock \showarticletitle{Transferable Unsupervised Outlier Detection Framework for Human Semantic Trajectories}. In \bibinfo{booktitle}{\emph{SIGSPATIAL'24 (To Appear, Research Track)}}.
\newblock


\bibitem[Zhang et~al\mbox{.}(2023)]%
        {zhang2023large}
\bibfield{author}{\bibinfo{person}{Zheng Zhang}, \bibinfo{person}{Hossein Amiri}, {et~al\mbox{.}}} \bibinfo{year}{2023}\natexlab{}.
\newblock \showarticletitle{Large Language Models for Spatial Trajectory Patterns Mining}.
\newblock \bibinfo{journal}{\emph{arXiv e-prints}} (\bibinfo{year}{2023}), \bibinfo{pages}{arXiv--2310}.
\newblock


\bibitem[Zheng et~al\mbox{.}(2011)]%
        {zheng2011geolife}
\bibfield{author}{\bibinfo{person}{Yu Zheng}, \bibinfo{person}{Hao Fu}, \bibinfo{person}{Xing Xie}, \bibinfo{person}{Wei-Ying Ma}, {and} \bibinfo{person}{Quannan Li}.} \bibinfo{year}{2011}\natexlab{}.
\newblock \bibinfo{booktitle}{\emph{GeoLife GPS trajectory dataset - User Guide} (\bibinfo{edition}{geolife gps trajectories 1.1} ed.)}.
\newblock
\urldef\tempurl%
\url{https://www.microsoft.com/en-us/research/publication/geolife-gps-trajectory-dataset-user-guide/}
\showURL{%
\tempurl}


\bibitem[Zhu et~al\mbox{.}(2024a)]%
        {zhu2024generic}
\bibfield{author}{\bibinfo{person}{Feng Zhu}, \bibinfo{person}{Cheng Chang}, \bibinfo{person}{Zhiheng Li}, \bibinfo{person}{Boqi Li}, {and} \bibinfo{person}{Li Li}.} \bibinfo{year}{2024}\natexlab{a}.
\newblock \showarticletitle{A generic optimization-based enhancement method for trajectory data: Two plus one}.
\newblock \bibinfo{journal}{\emph{Accident Analysis \& Prevention}}  \bibinfo{volume}{200} (\bibinfo{year}{2024}), \bibinfo{pages}{107532}.
\newblock


\bibitem[Zhu et~al\mbox{.}(2024b)]%
        {zhu2024synmob}
\bibfield{author}{\bibinfo{person}{Yuanshao Zhu}, \bibinfo{person}{Yongchao Ye}, \bibinfo{person}{Ying Wu}, \bibinfo{person}{Xiangyu Zhao}, {and} \bibinfo{person}{James Yu}.} \bibinfo{year}{2024}\natexlab{b}.
\newblock \showarticletitle{SynMob: Creating High-Fidelity Synthetic GPS Trajectory Dataset for Urban Mobility Analysis}.
\newblock \bibinfo{journal}{\emph{Advances in Neural Information Processing Systems}}  \bibinfo{volume}{36} (\bibinfo{year}{2024}).
\newblock


\bibitem[Z{\"u}fle et~al\mbox{.}(2023)]%
        {zufle2023urban}
\bibfield{author}{\bibinfo{person}{Andreas Z{\"u}fle}, \bibinfo{person}{Carola Wenk}, \bibinfo{person}{}, {et~al\mbox{.}}} \bibinfo{year}{2023}\natexlab{}.
\newblock \showarticletitle{Urban life: a model of people and places}.
\newblock \bibinfo{journal}{\emph{Computational and Mathematical Organization Theory}} \bibinfo{volume}{29}, \bibinfo{number}{1} (\bibinfo{year}{2023}), \bibinfo{pages}{20--51}.
\newblock


\end{thebibliography}

\end{document}